\documentclass[12pt]{article}
\usepackage[pdftex]{graphicx}
\usepackage{amssymb,amsmath}
\makeatletter

\@addtoreset{equation}{section}
\def\section{\@startsection {section}{1}{\z@}{-2.5ex plus -1ex minus
 -.2ex}{1.3ex plus .2ex}{\large\bf}}
\def\subsection{\@startsection{subsection}{2}{\z@}{-2.25ex plus%
 -1ex minus -.2ex}{0.5ex plus .2ex}{\bf}}

\advance \voffset by -0.8in
\advance \hoffset by -0.7in
\def\tJ{\tilde{J}}
\def\tP{\tilde{P}}
\def\glambda{\mathfrak{g}_\lambda}
\def\Glambda{G_\lambda}

\def\Ad{\mbox{Ad}}

\def\bn{{\mbox{\boldmath $n$}}}

\def\bpm{\begin{pmatrix}}
\def\epm{\end{pmatrix}}

\newcommand{\ZZ}{\mathbb{Z}}

\newcommand{\RR}{\mathbb{R}}
\newcommand{\CC}{\mathbb{C}}

\def\bea{\begin{eqnarray}}
\def\eea{\end{eqnarray}}

\newcommand{\m}{m}

\def\rbiprod{{\cdot\kern-.33em\triangleright\!\!\!<}}
\def\lbiprod{{>\!\!\!\triangleleft\kern-.33em\cdot}}

\begin{document}
\parskip 6pt
\parindent 0pt
\begin{flushright}
EMPG-11-11\\
\end{flushright}

\begin{center}

\baselineskip 28 pt 

{\Large \bf   Quantum  gravity  and  non-commutative spacetimes in  three dimensions: a unified approach}

\baselineskip 18 pt 

 { Bernd~J.~Schroers\footnote{\tt bernd@ma.hw.ac.uk} \\
Department of Mathematics and Maxwell Institute for Mathematical Sciences \\
 Heriot-Watt University \\
Edinburgh EH14 4AS, United Kingdom }

\vspace{0.5cm}
Talk given at `Geometry and Physics in Cracow', September 2010

\end{center}

\begin{abstract}
\noindent  These notes summarise a talk  surveying   the combinatorial or Hamiltonian quantisation of three dimensional gravity in the Chern-Simons formulation, with an emphasis on  the role of quantum groups and on the way the various physical constants ($c,G,\Lambda,\hbar)$  enter as deformation parameters. 
The classical situation is summarised,  where solutions  can be characterised in terms of  model spacetimes (which depend on $c$ and $\Lambda$)  together with global identifications via elements of the corresponding isometry groups.  The quantum theory  may be viewed as a deformation of this picture, with quantum groups replacing the local isometry groups, and non-commutative spacetimes replacing the classical model spacetimes. This point of view is  explained, and open issues are sketched. 

\end{abstract}


\section{Introduction and motivation}
\subsection{Historical remarks}  Giving a talk on three dimensional (3d) gravity at a meeting in Cracow  is like carrying coal to Newcastle: 
the beginnings of  the subject  are usually traced  back to  the paper  \cite{Staruszkiewicz} by Andrzej Staruszkiewicz, alumnus and later professor at the Jagellonian University in Cracow. Staruszkiewicz's paper, published in 1963,  is about classical 3d gravity and its special  features.  The subject of 3d {\em quantum} gravity started only  five years later  with the realisation by Ponzano and Regge  \cite{PonzanoRegge}  that  angular momentum theory plays an important role in this context.

Gravity in 3d   is now a large subject in its own right, which I can not possibly review here. However, in this introductory part  of the talk I will at least attempt to identify a few of the main themes and relate them to the approach followed here. Influential  papers by Deser, 't Hooft and Jackiw written in the 1980s \cite{DJtH,DeserJackiw1,DeserJackiw2,tHooftscattering} on classical and quantum scattering of particles  demonstrated the possibility of carrying out non-perturbative calculations of quantum scattering processes in 3d gravity. As we shall see, they also contain  indications of the relevance of the braid group in describing such processes. These indications are elaborated in the later literature, see for example \cite{Carlipscattering,BM,BMS},  and turn out to be closely related to the quantum group approach  pursued in this talk.  

The Chern-Simons formulation of 3d gravity, observed in \cite{AT} and elaborated in \cite{Witten}, establishes a connection between 3d gravity and a host of areas in mathematical physics, including  topological field theory, knot theory, the theory of Poisson-Lie groups and of quantum groups. Since this talk is based on the Chern-Simons approach, we will see  many of these connections.

The  early paper by Ponzano and Regge, mentioned above,  provides the foundation of the spin foam   approach to 3d quantum gravity.  This is perhaps the approach to 3d quantum gravity that contains the most  directly useful lessons for  4d quantum gravity. I will not discuss this approach in this talk, and shall not attempt to summarise the large literature  on it. However, it is worth pointing out that  there are close links with  Chern-Simons theory (spin foam state sums may be viewed as discretisation of the path integral)   and to quantum groups, see  \cite{TuraevViro} for an early paper and \cite{Foxton,BarrettNaishGuzman} for examples of  recent papers with many references. 

The possibility that non-commutative geometry is needed to describe spacetime at the quantum level has  long been a theme in quantum gravity research \cite{DFR}, see \cite{BianchiRovelli} for a recent discussion with some references. It is therefore interesting to ask if one can use the relatively tractable  3d situation to establish the role of non-commutative geometry  in quantum gravity in a mathematically convincing way. Early discussions of non-commutative spacetime coordinates appear   in  the paper  \cite{tHooftdiscrete}. Spacetime non-commutativity in 3d quantum gravity is  studied,  in different approaches, in \cite{MatWell,BatistaMajid,FreidelLivine,JoungMouradNoui}. Putting these approaches into one coherent picture is one of the objectives of this talk.  

Finally,  I should mention two  further important themes of 3d gravity research  which I will not  be able to touch on in this talk. One is the study of BTZ black holes, an introduction to which can be found in the book \cite{Carlipbook}.  The other is the relation to  3d  hyperbolic geometry, where the papers and books  \cite{Mess,Messcomment,BenedettiBonsante,Thurston}  may provide  good starting points. 

\subsection{Topological degrees of freedom and interactions in 3d gravity}
\label{intro}
The Einstein field equations (without cosmological constant and in units where the speed of light is 1)
\[
 R_{ab}-\frac 1 2 R g_{ab} = -8\pi GT_{ab}
\]

determine the Ricci tensor of a spacetime in terms of the energy momentum tensor. In spacetime dimensions  greater than three, the  Ricci tensor does not fix the Riemann tensor and it is possible to have metrically non-trivial (i.e. curved)  spacetimes satisfying the vacuum ($T_{ab}=0$) field equations.   In three spacetime dimensions, this is not possible. The Ricci tensor determines the Riemann tensor and, as a result, the only  vacuum solutions of the Einstein equations  with  vanishing cosmological constant are  flat \cite{Carlipbook}. This result simplifies Einstein's theory of gravity in 3d dramatically, but does not render it trivial. There are non-trivial solutions of the Einstein equations in the presence of matter, and, if the topology of the three-dimensional manifold representing the universe is non-trivial, there may be vacuum solutions which, though flat,  have  non-trivial holonomy. These observations are often summarised in the slogan that in 3d gravity there are no gravitational waves 
but that the theory has  topological degrees of freedom. 

The simplest solution of the Einstein equations illustrating the previous paragraph is the spacetime surrounding a point-particle. The energy-momentum tensor is a Dirac delta-function with support on the world line of the particle.The metric solving the  field equations is flat away from the world line and is singular on the world line. More precisely it is a direct product of a cone (space) and $\RR$ (representing time) \cite{Carlipbook}. The line element, in terms of polar coordinates $(r,\phi)$, with $r>0$,  and a time coordinate $t$  is simply
\bea
\label{pointsolution}
 ds^2 = c^2dt^2 -dr^2 - r^2 d\phi^2.
\eea
However,  the range of $\phi$ is $[0,2\pi-\mu)$, where the parameter $\mu$ is related to the particle's mass $m$ and to Newton's constant $G$ via
\[
 \mu = 8\pi G m.
\]
In three dimensions , the physical dimension of $G$ is that of an inverse mass so that $\mu$ is a dimensionless, angular parameter. The effect of a particle on the geometry of spacetimes is, then, to cut out a wedge of size $\mu$ from the spacetime surrounding the particle's world line. 

\begin{figure}[!ht]
\centering
\includegraphics[width=5truecm]{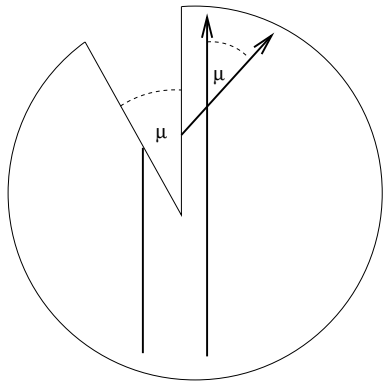}
\caption{Geodesics in the space surrounding a conical singularity with deficit angle $\mu$}
\label{conedeflection}
\end{figure}

It is instructive to consider the effect of the geometry \eqref{pointsolution} on light  test particles. Such particles travel on geodesics, which are simply straight lines on the cone after it has been cut open. It is easy to check that geodesics passing the particle of mass $m$ on one side are deflecting relative to particles who pass it on the other side by the angle $\mu$ (in the coordinate system $(t,r,\phi)$). This relative deflection is illustrated in Fig.~\ref{conedeflection} and is independent of the distance of closest approach  between the heavy particle of mass $m$ and the test particles (impact parameter). The interaction is topological in the sense that it only depends on whether the test particle passes on the left or the right of the heavy particle, and not on the  relative distance. This kind of interaction is familiar from the Aharonov-Bohm interaction between electrons and a magnetic flux, and  this analogy can be made precise: both interactions can be related to the braiding of the world lines of the interacting particles \cite{BMS}.

\subsection{Physical constants entering 3d quantum gravity}

The four physical constant entering 3d  quantum gravity are the speed of light $c$, Newton's constant $G$,  Planck's constant $\hbar$ and the cosmological constant $\Lambda$. From these, we can form two length constants (remembering that the dimension of $G$ is an inverse mass), namely
\bea
\label{constants}
 \text{Planck length} \;\;  \ell_P=\frac{\hbar G}{c}, \quad  \text{Cosmological length scale} \;\;\ell_C = \frac{1}{\sqrt{\Lambda}}.
\eea
In this talk we will deal with both Lorentzian and Euclidean gravity, and we parametrise Euclidean and Lorentzian metrics in a unified fashion by allowing $c^2<0$  in the Euclidean situation. As a result, both the length parameters in \eqref{constants} may be imaginary, depending on the sign of $c^2$ and $\Lambda$.  
From the ratio of the two length parameters we can form a dimensionless quantity. We define the deformation parameter
\bea
\label{qpara}
q= e^{-\frac{\hbar G\sqrt{\Lambda}}{c}}, 
\eea
which may take values on the real line or the unit circle in the complex plane.

It is useful to clarify the role played by the various constants in 3d gravity in general terms  at this stage. The observation  of the previous section that, in the absence of matter, solutions of the Einstein equations are locally flat generalises  in the presence of a cosmological constant to the statement that vacuum solutions are locally isometric to model space times, which depend on the parameters $c$ and $\Lambda$. For Lorentzian gravity with vanishing cosmological constant, for example, the model spacetime is Minkowski space while for Euclidean gravity with positive cosmological constant it is the four-sphere with the round metric. The isometry groups of the model spacetimes  inherit a dependence on $c$ and $\Lambda$. In the examples above they are, respectively, the Poincar\'e group in 3d and the 4d rotation group $SO(4)$.  
Newton's constant $G$ enters when one studies  the dynamics of spacetime  and plays the role of a parameter in the Poisson structure and that of a coupling constant to matter. Finally, $\hbar$ enters in the quantisation and the dimensionless parameter $q$ in \eqref{qpara}, combining all four constant, controls the quantum theory when all the constants  $1/c,G, \Lambda,\hbar$ are non-zero.

\subsection{Motivation and outline of the talk}
The goal of this talk is give a unified account of  aspects of classical and quantum gravity in 3d, in which the physical parameters of the previous section enter as deformation parameters.  Our account of  classical gravity is  based on the formulation of 3d gravity as a Chern-Simons gauge theory, where  the local isometry groups   play the role of the gauge groups.  As well shall see, the parameters $c$ and $\Lambda$ enter in this description via the structure constants of the Lie algebra of the gauge group, while the parameter $G$ enters via the inner product (or trace) on the Lie algebra which is used in  the Chern-Simons action. We sketch the description of the phase space of 3d gravity as the moduli space of flat connections,  and review the description of its Poisson structure in a formulation, due to Fock and Rosly \cite{FockRosly}, which makes essential use of classical $r$-matrices. 

The description of the Poisson structure in terms of $r$-matrices is tailor-made for  the quantisation via the combinatorial or Hamiltonian scheme pioneered in \cite{AGSI},
\cite{AGSII} and \cite{AS}. In this scheme, the quantisation is controlled by quantum groups which are  deformations of the  local isometry groups of the model spacetimes, with deformation parameters $G$ and $\hbar$ in addition to $c$ and $\Lambda$. These quantum groups naturally act on non-commutative spaces,  which one may interpret as deformations of the classical model spacetimes. This framework thus provides a concrete mathematical setting for exploring the proposal  that, in quantum gravity, spacetime should be mathematically modelled in terms of non-commutative geometry. We end our talk with an evaluation of the successes and limitations of this approach to 3d quantum gravity.


\section{Model spacetimes and isometry groups}

\begin{figure}[!ht]
\centering
\includegraphics[width=5.5truecm]{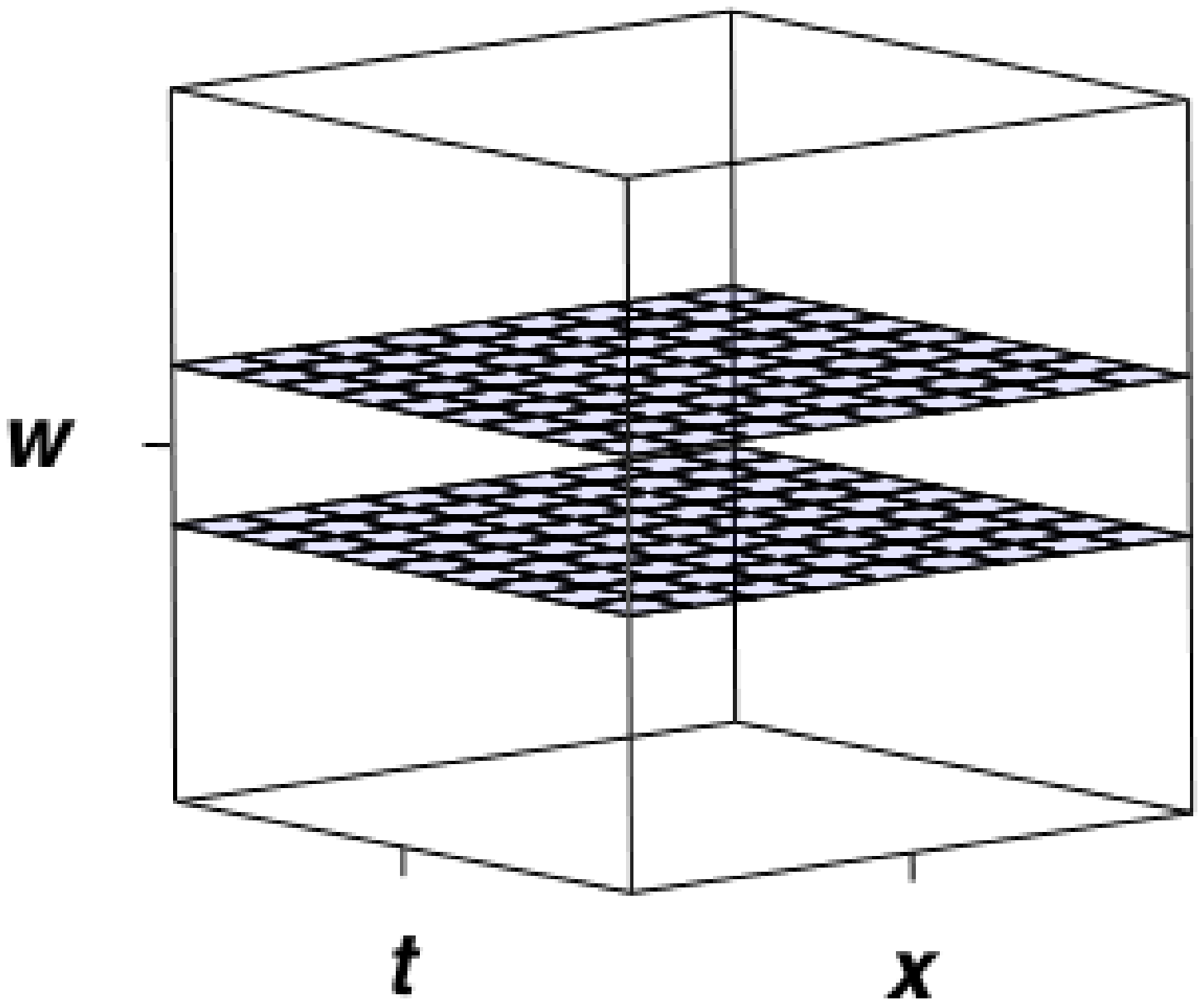}\\
\vspace{-.4cm}
\includegraphics[width=5.5truecm]{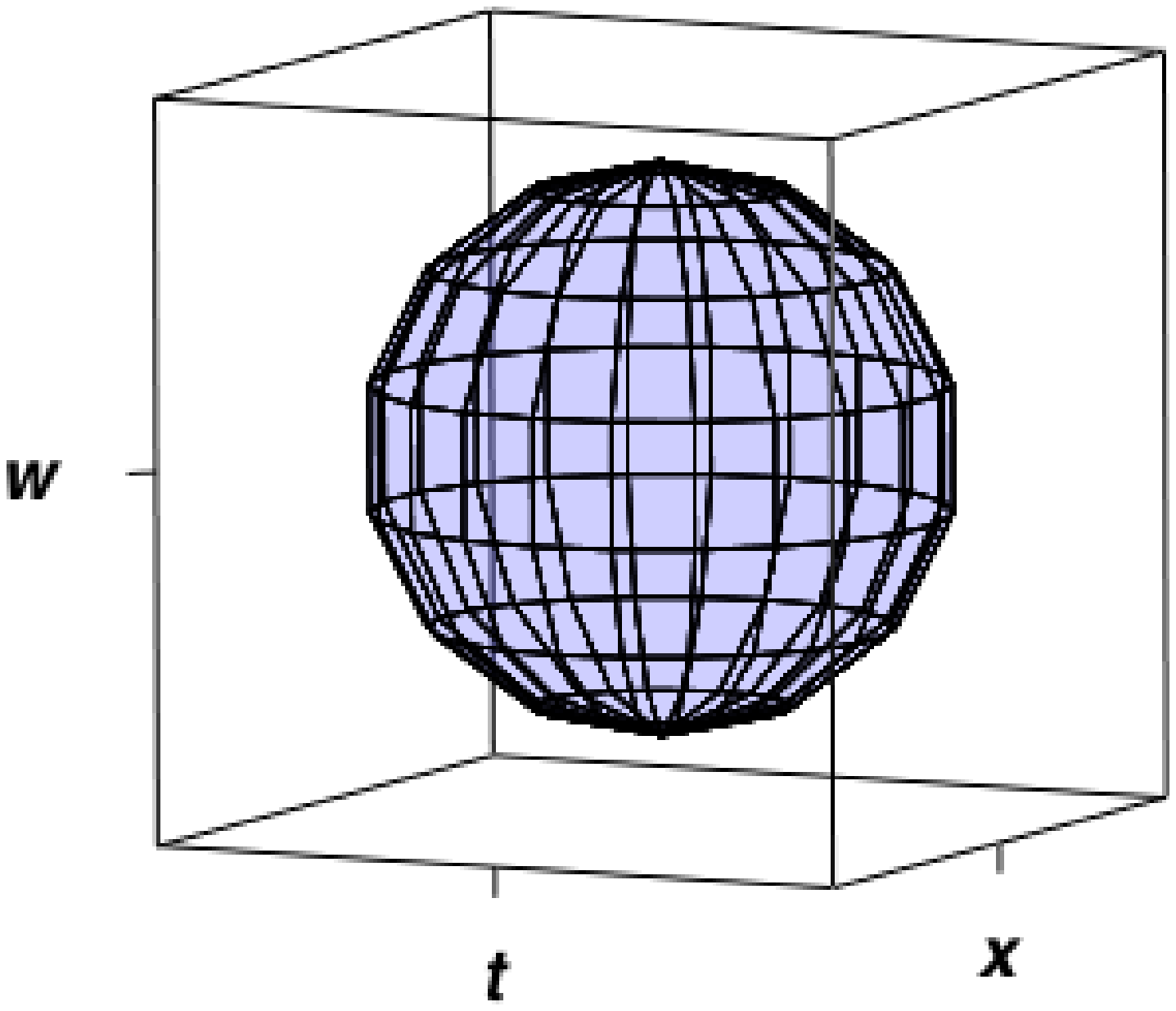}
\includegraphics[width=5.5truecm]{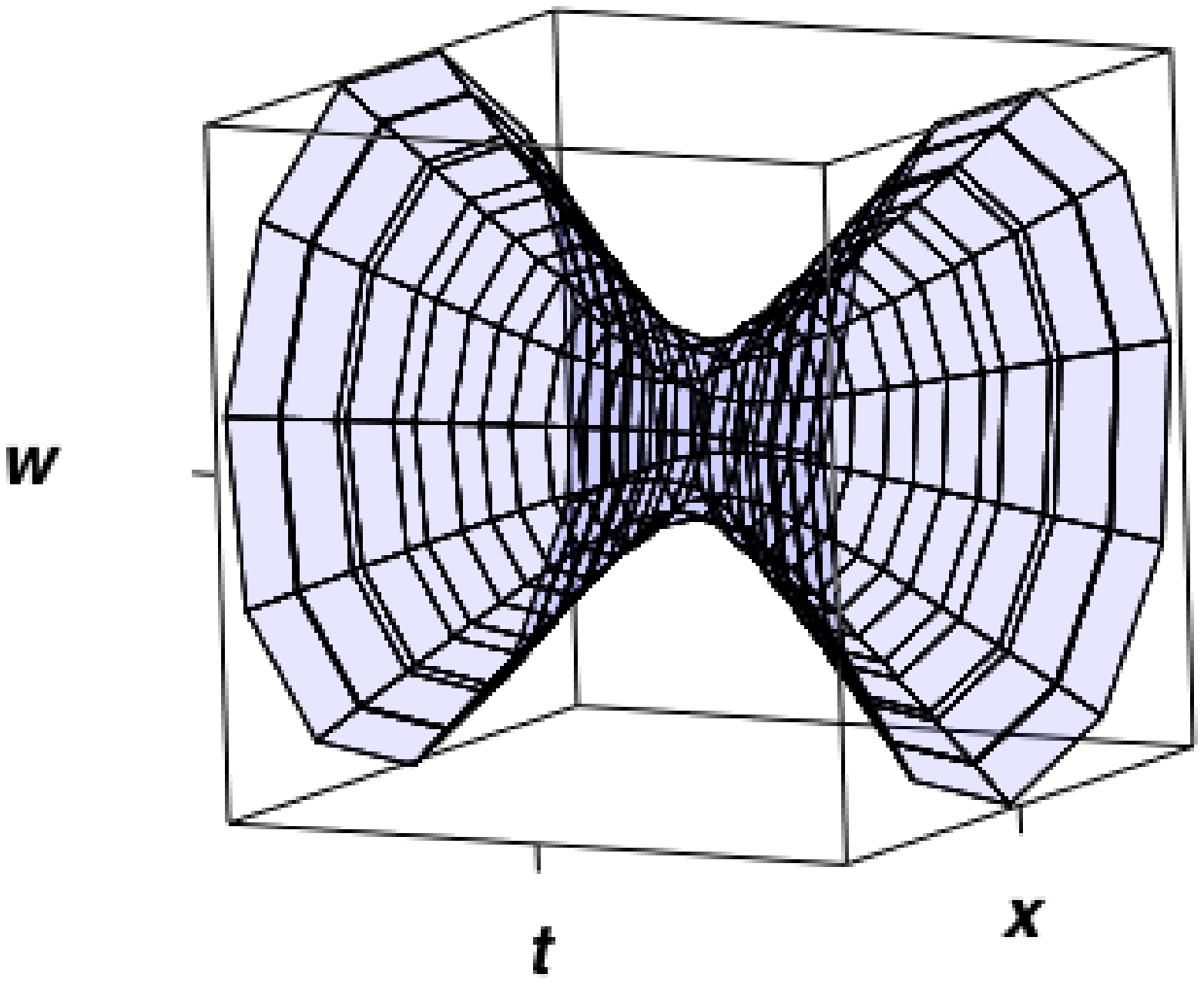} 
\\
\includegraphics[width=5truecm]{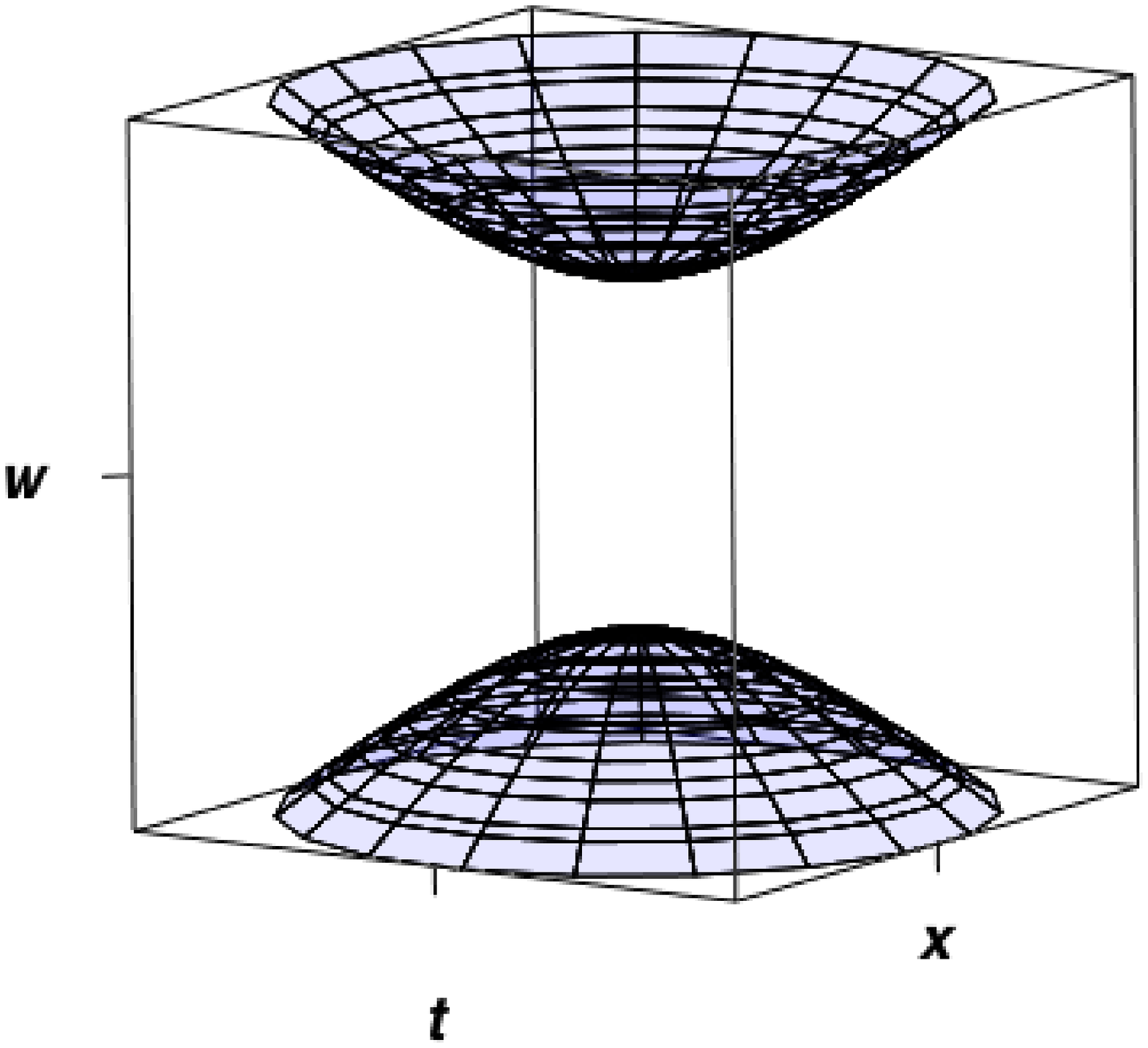} \;\;\;\;\;\; 
\includegraphics[width=5.5truecm]{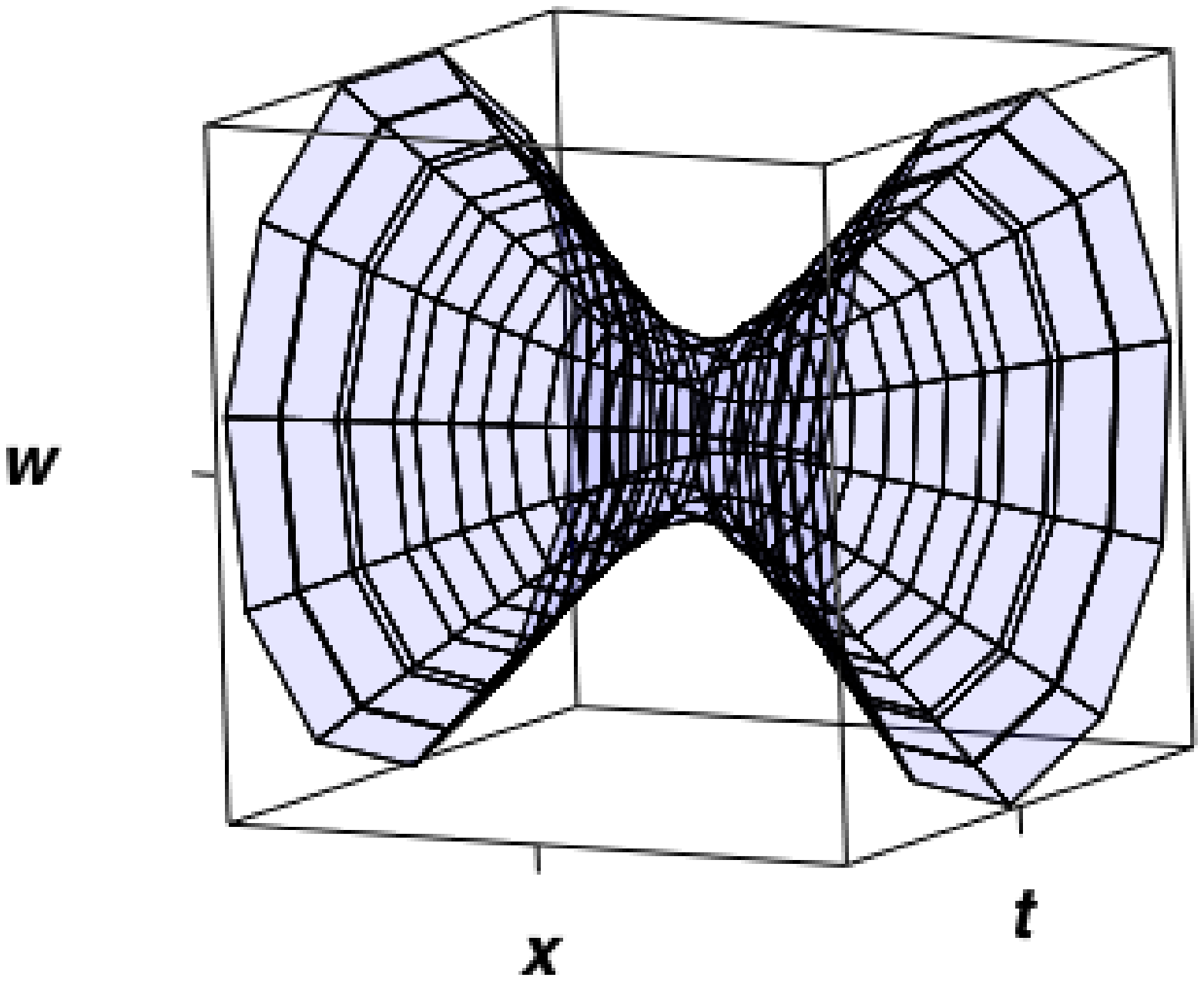} 
\caption{(Double covers of) Model spacetimes for 3d gravity, shown as 2d models embedded in a 3d auxiliary space
 with coordinates $(t,x,w)$ according to \eqref{hypersur} (the second spatial coordinate $y$ is suppressed).
Euclidean and Minkowski space at the top, spherical and de Sitter space in the middle, hyperbolic and Anti-de Sitter space at the bottom}
\label{modelspaces}
\end{figure}

The following treatment of the model spacetimes  follows closely that in  \cite{PapageorgiouSchroers1}. 
We use Roman letters $a,b,c \ldots$ for  3d spacetimes indices, with range for  $\{0,1,2\}$ (in both the Euclidean and Lorentzian case).
The model spacetimes arising in 3d gravity can be described in a simple an unified fashion in  terms of the metric
\bea
\label{4dmetric}
g_{\mu\nu}=\text{diag}\left(-c^2,1,1,\frac 1 \Lambda\right)
\eea
in an auxiliary $\RR^4$. Here  we use Greek indices  for the range $\{0,1,2,3\}$.
The model spacetimes can be realised as  embedded hypersurfaces  via
\bea 
\label{hypersur}
H_{c,\Lambda}=
\left\{(t,x,y, w)\in \RR^4| -c^2t^2 + x^2 + y^2 + \frac {1}{\Lambda} w^2 = \frac{1}{\Lambda}\right\} .
\eea
This two-parameter family includes the  three-sphere $S^3$ ($c^2<0$, $ \Lambda >0$),  doubles covers of hyperbolic space $H^3$ ($c^2 <0$,  $\Lambda <0$), de Sitter space dS$^3$ ($c^2>0$, $ \Lambda >0$) and anti-de Sitter space AdS$^3$ ($c^2>0$, $\Lambda <0$). Double covers of 
Euclidean  space $E^3$  and Minkowski $M^3$ space arise in the limit  $\Lambda\rightarrow 0$, which one should take {\em after} multiplying the defining equation in \eqref{hypersur} by $\Lambda$.  In Fig.~\ref{modelspaces} we show the  embedded model spacetimes  (with one spatial dimension suppressed).

In order to be able to take the limit $\Lambda \rightarrow 0$  for the associated isometry groups it is best to work with the inverse metric
\bea
\label{invmet}
g^{\mu\nu}=\text{diag}\left(-\frac{1}{c^2},1,1,\Lambda\right).
\eea
 The Lie algebra generators of the isometry groups of \eqref{invmet} can conveniently be defined in terms  of the Clifford algebra associated to \eqref{invmet} \cite{PapageorgiouSchroers1}.
Thus we define generators $\gamma^\mu$ via
\bea
\{\gamma^\mu,\gamma^\nu\}=-2g^{\mu\nu},
\eea
so that the six  Lie algebra generators are given by
\bea
M^{\mu\nu}=\frac 1 4 [\gamma^\mu,\gamma^\nu].
\eea
They have the commutation relations
\bea
\label{masterlie}
[ M^{\kappa\lambda}, M^{\mu\nu} ] = 
 g^{\kappa\mu}  M^{\lambda\nu} + g^{\lambda\nu} M^{\kappa\mu}- g^{\kappa\nu} 
 M^{\lambda\mu}-g^{\lambda\mu} M^{\kappa\nu}.
 \eea
The advantage of the Clifford algebra approach is that one  can immediately write down two naturally defined  invariant bilinear forms. One, denoted $\langle\cdot , \cdot \rangle$  is defined by carrying out the Clifford multiplication and projecting onto the  invariant, central element  $\gamma^5=\gamma^0\gamma^1\gamma^2\gamma^3$.
Multplying by $-4$ for later convenience, the resulting inner product is non-zero whenever the indices on the basis vectors are complementary, for example
\[
 \langle M^{12},M^{03}\rangle=-1, \quad \langle M^{12},M^{01}\rangle=0 \quad \text{etc.}
\]
Another bilinear form $(\cdot, \cdot)$ is obtained by carrying out the Clifford multiplication and projecting onto the identity. Again rescaling by $-4$ for convenience we have
a  non-zero answer  whenever the indices on the basis vectors match:
\[
 (M^{12},M^{12})=1,\quad (M^{01},M^{01})=-\frac{1}{c^2},
 \quad (M^{13},M^{13})=\Lambda \quad \text{etc.}
\]
As we shall see shortly, this is the Killing form on the Lie algebra

 We now express the above generators in more conventional 3d notation. For this purpose we  define the three-dimensional  totally antisymmetric  tensor with downstairs indices via
 $\epsilon_{012}=1$. 
 Then  we define the rotation generator $\tJ_0$, the boost generators $\tJ_1,\tJ_2$ and translation generators $\tP_a$ via
 \bea
\label{liegens}
 \tJ_a=\frac  1 2 \epsilon_{abc}M^{bc},\quad\quad \tP_a=g_{ab}M^{b3},
  \eea
  where we used the spacetime part of the 4d metric $g_{\mu\nu}$ \eqref{4dmetric} to lower indices, and refer to 
  \cite{PapageorgiouSchroers1} for a discussion of physical dimensions and interpretation of these generators (which are denoted by the same letters, but without tilde there). 
  The Lie algebra brackets are now
    \bea
  \label{t21alg}
  [\tJ_a,\tJ_b]=\epsilon_{abc}J^c, \quad [\tJ_a,\tP_b]=\epsilon_{abc} \tP^{ c},\quad [\tP_a,\tP_b]=-c^2\Lambda
  \epsilon_{abc}\tJ^c,
  \eea
  with indices raised via the inverse  metric $g^{ab}$.
  The combination $-c^2\Lambda$ which occurs in the Lie brackets plays an important role in what follows, and we introduce
\bea
\label{algebraicconstant}
\lambda =-c^2\Lambda.
\eea
 The bilinear form $(\cdot, \cdot)$, already advertised as the Killing form,  is 
   \bea
\label{nongravin}
  (\tJ_a,\tJ_b)=\kappa_{ab},\quad (\tP_a,\tP_b)=\lambda\kappa_{ab}, 
  \eea
where 
\bea
\kappa_{ab}=-\frac{1}{c^2} \;g_{ab} = \text{diag}\left(1, -\frac{1}{c^2} ,-\frac{1}{c^2}\right).
 \eea
The metric $\kappa_{ab}$ is the most natural one on the Lie algebra  $so(3)$ respectively   $so(2,1)$ spanned by $\tJ_0,\tJ_1$ and $\tJ_2$. Note that it differs  from the spacetime metric $g_{ab}$, but that it has the right physical dimensions and that imaginary  $c$ gives the usual Euclidean metric, as required.

It is one of the coincidences of 3d that spacetime and the Lie algebra of  rotations and/or boosts are both three-dimensional. Both are equipped with Euclidean respectively Lorentzian metrics, but our derivation shows that, in a physically natural normalisation and  construction,  the spacetime and Lie algebra metrics come out differently. This is potentially confusing in calculations where indices are raised and contracted with these metrics, and most papers on 3d gravity use conventions where the two kinds of metrics coincide. We can achieve this by switching from the physical Lie algebra basis  used thus far to a geometrical basis  according to 
\begin{align}
\tJ_0 &\rightarrow J_{0}=-\frac{|c|^2}{c^2} \tJ_0,\quad  \tJ_1 \rightarrow J_1 = |c| \tJ_1, \quad \tJ_2  \rightarrow J_2 = |c| \tJ_2, \nonumber \\
\tP_0 &\rightarrow  P_{0}=-\frac{|c|^2}{c^2} \tP_0,\quad  \tP_1 \rightarrow P_1 = |c| \tP_1, \quad \tP_2  \rightarrow P_2 = |c| \tP_2.
\end{align}
In this geometrical basis, all the generators $J_a$ are dimensionless, and all the translation generators $P_a$ have the dimension of inverse time. One checks that the Killing metric now takes the form
\bea
\label{3dmetric}
(J_a,J_b)= \eta_{ab}:=\text{diag}\left(1,-\frac{|c|^2}{c^2}, -\frac{|c|^2}{c^2}\right),
 \eea
 which is diag$(1,1,1)$ in the Euclidean and diag$(1,-1,-1)$ in the Lorentzian case. Moreover, the Lie 
 brackets take the {\em same} form as in \eqref{t21alg}, 
 \bea
  \label{21alg}
  [J_a,J_b]=\epsilon_{abc}J^c, \quad [J_a,P_b]=\epsilon_{abc} P^{ c},\quad [P_a,P_b]=\lambda
  \epsilon_{abc}J^c, 
  \eea
but all indices are now raised with the Lie algebra metric $\eta_{ab}$. This is convenient and we shall work in this basis for the remainder of this talk.  We denote the Lie algebra with these  brackets   by $\glambda$.  The conventions regarding the  metric then agree with 
  \cite{MeusburgerSchroers6}, but the convention regarding the naming of $\lambda$  agrees with \cite{Witten} and differs from \cite{MeusburgerSchroers6}, where  $\Lambda$ was used for what we call $\lambda$ now. Conventions regarding the naming of the cosmological constant and the combination \eqref{algebraicconstant}  differ in the literature, and the reader will need to take good care when comparing results from different sources.

 The other  bilinear form  introduced in the Clifford language gives the following non-zero pairings 
  \bea
\label{3dgravin}
  \langle J_a, P_b\rangle =c^2\eta_{ab}.
  \eea
 This   pairing is non-degenerate for any value of $\lambda$ and is crucial for  the Chern-Simons formulation of 3d gravity, as we shall see.  
   
In Table~\ref{isogroups} we list Lie groups whose Lie algebras are \eqref{21alg}.  We have used the isomorphisms $SU(2)/\ZZ_2 = SO(3)$
and $SL(2,\RR)/\ZZ_2= SO(2,1)_0$, the identity component of $SO(2,1)$. The isometry groups are determined by their Lie algebras only up to coverings, and our choice in Table~\ref{isogroups} is  one of  convenience. In the following, we  write $\Glambda$ for this family of Lie groups. 
 
\begin{table}[h!]
\begin{center}
\begin{tabular}{|c|c|c|}
\hline
  &    &   \\
Cos. constant & Euclidean ($c^2<0$) & Lorentzian ($c^2>0$) \\
 &    &   \\
\hline
  &    &   \\
$\Lambda =0$  & $SU(2)\ltimes \RR^3$ & $SL(2,\RR) \ltimes \RR^3$ \\
 &    &   \\
\hline
  &    &   \\
$\Lambda> 0$ &
$ SU(2)\times SU(2)$&
$SL(2,\CC)$ \\
  &    &   \\
\hline
  &    &   \\
$ \Lambda< 0$ & $ SL(2,\CC) $&
$SL(2,\RR)\times SL(2,\RR) $\\
&    &   \\
\hline
\end{tabular}
\caption{Local isometry groups in 3d gravity}
\label{isogroups}
\end{center}
\end{table}

\section{The Chern-Simons formulation of 3d gravity}

In Cartan's approach to  Riemannian geometry  \cite{Sharpe} the fundamental geometrical object is a  connection which combines an orthonormal frame field (or  vielbein) $e_a$  and the spin connection $\omega_{ab}$  on the orthonormal frame bundle into  the so-called Cartan connection. Concretely,  in the case of 3d geometry, we  combine the dreibein with the translation generators $P_a$ of \eqref{liegens}  and  the local connection one-forms  $\omega^a = \frac 1 2 \epsilon^{abc}\omega_{bc}$ with the rotation and/or Lorentz generators $J^a$ into the local one-form 
 \bea
\label{Cartan}
 A=e_aP^a + \omega_aJ^a,
 \eea
taking values in the Lie algebra $\glambda$.  
The curvature 
\bea
\label{Cartancurv}
F_A=dA+\frac 1 2[A\wedge A] =R+C+T
\eea
  of the Cartan connection   combines
the Riemann curvature  of the spin connection $\omega= \omega_a J^a$,
\[
R= d\omega + \frac 1 2 [\omega\wedge\omega],
\]
a cosmological term
\[ C=\frac \lambda 2\epsilon^{abc} e_a\wedge e_b J_c,
 \]
and the  torsion 
\[
T=(de^c +\epsilon^{abc}\omega_a\wedge e_b) P_c.
\]

In the Cartan  approach to general relativity (in any dimension),  the Einstein-Hilbert action is expressed in terms of the vielbein and the  connection, which  are treated as independent variables. The action is  called the Palatini action when interpreted in this way.   In this approach,  the condition of vanishing torsion (in the absence of spin sources)  follows as a variational equation rather than as an a priori condition. 
It turns out that, in three dimensions,    the Einstein-Hilbert (or Palatini) action is simply the Chern-Simons action for the Cartan connection \eqref{Cartan}, with the bilinear form \eqref{3dgravin} used as an inner product \cite{AT,Witten}. However, beyond the equality of the actions,  the  relationship between the Chern-Simons formulation and the Einstein formulation of 3d gravity is subtle:  non-invertible dreibeins $e_a$  may occur in the Chern-Simons formulation but  are ruled out in the Einstein approach, based on metrics. This changes the nature of gauge orbits in the two   cases, so that the physical phase spaces are, in general, different. This was pointed out in a 1+1 dimensional context in \cite{SchS} and was demonstrated in an explicit example involving four particles  in 3d gravity in \cite{Matschull2}. Our approach to 3d gravity in the remainder of this talk is based on the Chern-Simons formulation. 

We discuss the Chern-Simons action in terms of the general   bilinear form 
\bea
\label{genform}
 (\cdot ,\cdot )_{\alpha\beta} = \alpha\langle\cdot, \cdot\rangle + \beta (\cdot ,\cdot )
\eea
on the Lie algebra $\glambda$. This form  is non-degenerate  iff  \cite{MeusburgerSchroers7}
\bea
\label{nondegcond} 
\alpha^2  - \lambda \beta^2\neq 0,
\eea  
and the associated action 
\begin{align}
\label{genact}
I_{\alpha\beta}(A)&=\int_M (A\wedge dA)_{\alpha\beta}+ \frac 1 3 (A\wedge[A,A])_{\alpha\beta}
\nonumber \\
  &=\alpha \int_M\left(2  e^a\wedge R_a  + \frac \lambda 3 \epsilon_{abc}
  e^a\wedge e^b\wedge e^c \right)  \nonumber   \\
  &+\beta \int_M\left(\omega^a\wedge d\omega_a +\frac{1}{3}\epsilon_{abc}\omega^a
 \wedge \omega^b\wedge\omega^c + \lambda e^a\wedge T_a\right),
\end{align}
contains the  gravitational action  (the terms proportional to $\alpha$), the Chern-Simons action for the spin connection and  additional terms involving torsion.  
 This general action was first considered by Mielke and Baekler \cite{MielkeBaekler} and recently revisited in  \cite{BonzomLivine}, where the analogy between the terms proportional to $\beta$ and the Immirzi term in 4d was stressed.  The variational equations which follow from the  general action \eqref{genact} are simply the flatness condition for the Cartan connection, i.e. the vanishing of \eqref{Cartancurv}, provided the form \eqref{genform} is non-degenerate. This appears to imply that the family of actions \eqref{genact} leads to equivalent physics provided  the condition \eqref{nondegcond} holds. However, as argued in \cite{MeusburgerSchroers7}, the induced canonical  structure of the phase space does depend on the ratio of $\alpha$ and $\beta$.  Since we are only interested in the Chern-Simons formulation of  3d gravity here, we set 
\bea
\label{alphachoice}
\alpha =\frac{1}{16\pi G}, \qquad \beta = 0 
\eea
from now onwards. 

The gauge formulation of 3d gravity can easily and naturally be extended to include  minimal coupling between the gauge field and point particles. This was first discussed in in detail in \cite{SousaGerbert} and is  reviewed in our notation in  \cite{MeusburgerSchroers7}, where the dependence of the coupling on the parameters $\alpha$ and $\beta$ is also discussed. We are not able to discuss the coupling to particles, the  Poisson structure and the  division by gauge equivalence in the space available here.  Instead, we summarise the results   in the next section, and motivate them  in general, geometric terms.

\section{Classical $r$-matrices and Poisson brackets on the space of holonomies}

 Having established that, in the  Chern-Simons  formulation, 
classical solutions of the field equations are  flat $\Glambda$-connections, we can characterise the phase space of 3d gravity on a manifold $M^3$  in the Chern-Simons formulation as the space of flat $ \Glambda$-connections on  $M^3$, modulo gauge transformations.  In order to make this precise and concrete, we 
consider  3d universes of topology $M^3=\RR\times S$,  where $S$ is a two-dimensional manifold representing space. Then one can show  \cite{Witten} that the phase space is the  moduli space of flat $G_\lambda$-connections on $S$ (i.e. the space of flat $G_\lambda$-connections moduli gauge transformations),  equipped with the Atiyah-Bott symplectic structure \cite{AtiyahBott,Atiyah} , which is  defined in terms of the bilinear form used in the Chern-Simons action. With  the choice \eqref{alphachoice} this bilinear form is 
\bea
\label{gravinprod}
\frac{1}{16 \pi G} \langle \cdot, \cdot \rangle.
\eea
Therefore, in the Chern-Simons formulation, and assuming the factorisation $M^3=\RR\times S$, the task of constructing a theory of quantum gravity amounts to quantising the moduli space of flat $\Glambda$-connections on $S$, with a symplectic structure induced by \eqref{gravinprod}.

Despite the elegance and generality of this result,  a precise mathematical description  of this moduli space and  a rigorous quantisation remains a difficult task. In the case where $S$ is  a compact surface of genus $g\geq 2$, the moduli space can characterised in terms of the moduli space $\mathcal{A}_S$ of flat $SU(2)$  connections in the Euclidean case and in terms of Teichm\"uller Space $\mathcal{T}_S$  (a component of the moduli space of flat $SL(2,\RR)$ connections) in the Lorentzian case. In Table~\ref{phasesummary}  we reproduce  a summary of the results given in   \cite{KrasnovSchlenker}, where further references can be found.  The results  in the Lorentzian case are due to 
 \cite{Mess,Messcomment}. 
\begin{table}[h!]
\begin{center}
\begin{tabular}{|c|c|c|}
\hline
  &    &   \\
Cos. constant   & Euclidean ($c^2<0$) & Lorentzian ($c^2>0$) \\
  &    &   \\
\hline
 &    &   \\
$ \Lambda=0$  & $T^*\mathcal{A}_S$ & $T^*\mathcal{T}_S$ \\
 &    &   \\
\hline
  &    &   \\
$\Lambda > 0$ &
$\mathcal{A}_S\times \mathcal{A}_S$&
$\mathcal{T}_S\times \mathcal{T}_S\sim T^*\mathcal{T}_S$ \\
  &    &   \\
\hline
  &    &   \\
$ \Lambda < 0$ & $\mathcal{T}_S\times \mathcal{T}_S\subset T^*\mathcal{T}_S$&
$\mathcal{T}_S\times \mathcal{T}_S\sim T^*\mathcal{T}_S$\\
 &    &   \\
\hline
\end{tabular}
\caption{Phase space of 3d gravity for universes of the form $\RR\times S$, with $S$ compact and genus $\geq 2$ (quoted from \cite{KrasnovSchlenker}).}
\label{phasesummary}
\end{center}
\end{table}

For each of the  symplectic manifolds   in  the table, one may in principle attempt a quantisation and subsequent interpretation in terms of 3d quantum gravity. 
In this talk I summarise a description of the moduli space and its Poisson structure which is closely based the parametrisation  in terms of $\Glambda$-valued holonomies, and  which uses a concrete and unified description of the Poisson structure, which is tailor-made for quantisation. The idea for this description  is due to Fock and Rosly \cite{FockRosly}.  It is the foundation of the combinatorial  or Hamiltonian quantisation programme for Chern-Simons theory, described in \cite{AGSI,AGSII,AS}. 

Fock and Rosly's description of the phase space starts with the  observation that flat  connections on a manifold are characterised by their holonomies along non-contractible paths. The  moduli space of flat connections  on a surface  $S$  can thus be parametrised by the set of holonomies along closed paths which generate the fundamental group of $S$, modulo gauge transformations at the common starting and end point of those paths. So far we have assumed that $S$ is a compact manifold without boundary, but in the Fock and Rosly description it is easy to include punctures decorated with co-adjoint orbits  of $\Glambda$.  This is desirable in the context of 3d gravity, since  a co-adjoint orbit  of  $\Glambda$  physically correspond to  the phase space of a point particle, and the  `decoration' of a puncture with a co-adjoint orbit is precisely the effect of minimal coupling between the  Cartan connection  \eqref{Cartan} and the point particle's degrees of freedom. Moreover, this minimal coupling correctly reproduces the gravitational coupling between a point particle and the gravitational field, with momentum acting as a source of curvature and spin acting as a source for torsion. For details we refer the reader to   the papers \cite{SousaGerbert, MeusburgerSchroers7} and for a  relatively brief but pedagogical account to the talk  \cite{sissatalk}.

\begin{figure}[!ht]
\centering
\includegraphics[width=6truecm]{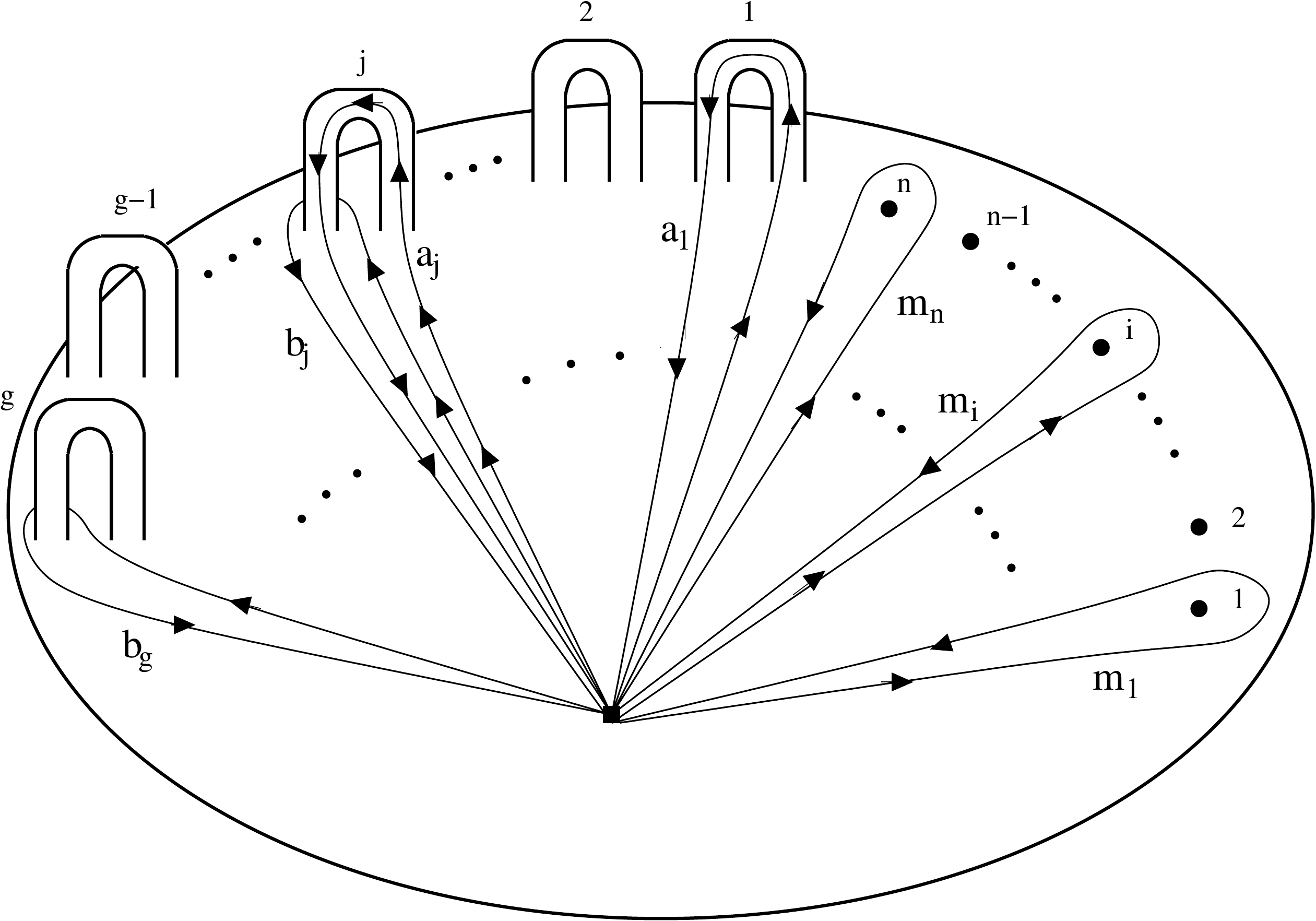}
\label{fundamentalgroup}
\caption{Generators of the fundamental group of a compact surface with punctures}
\end{figure}

The effect of the minimal coupling to co-adjoint orbits on the holonomies can be summarised as follows. Using  the inner product \eqref{gravinprod}, co-adjoint orbits can be written as adjoint orbits. For particles with mass $m$ and spin $s$, 
these orbits are of the form
\[
 O_{ms}=\{g(-\mu J_0 - \sigma P_0)g^{-1} | g \in \Glambda\},
\]
where
\[
 \mu =8\pi G\m, \quad \sigma =8\pi G s.
\]
Decorating a puncture on $S$ with  such an orbit forces the holonomy around the puncture to lie in the conjugacy class
\[
 {\cal C}_{\mu \sigma}=\{g(\exp(-\mu J_0 - \sigma P_0))g^{-1} | g \in \Glambda\}.
\]
For a genus $g$ surface $S$ with $n$ punctures   and orbit labels $\mu_i,\sigma_i$, $i=1\ldots n$,  a set of generators of the fundamental groups is shown in Fig.~\ref{fundamentalgroup}. The moduli space of flat $\Glambda$-connections can be written in terms of the extended phase space 
\bea
\label{extended}
\tilde {\cal P} = \Glambda^{2g}\times {\cal C}_{\mu_n \sigma_n}\times \ldots {\cal C}_{\mu_1 \sigma_1},
\eea
by  imposing the condition that  a suitable composition of the generating loops is contractible (and hence has trivial holonomy), and by dividing by conjugation at the base point:
\begin{align}
 {\cal P} & = \{(A_g,B_g,\ldots, A_1,B_1, M_n,\ldots M_1)\in \tilde {\cal P}| \nonumber \\
   & \quad [A_g,B_g^{-1}]\ldots[ A_1,B_1^{-1}]M_n\ldots M_1=1\}/\mbox{conjugation}.
\label{truephase}
\end{align}

The trick introduced by Fock and Rosly is to  define a (symplectic)  Poisson structure on the extended phase space $\tilde {\cal P}$  \eqref{extended} in such a way that the $\Glambda$-conjugation  action  on $\tilde {\cal P} $ is symplectic and that the symplectic quotient by it gives $\cal P$ with the Atiyah-Bott symplectic structure. The Poisson structure on  $\tilde {\cal P}$ is defined in terms of a classical $r$-matrix, i.e. an element $r\in \glambda\otimes \glambda$ which  satisfies the classical Yang-Baxter equation (CYBE)
\bea
\label{cybe}
[r_{12},r_{13}]+[r_{12},r_{23}]+[r_{13},r_{23}]=0, 
\eea
where we have used standard notation, explained,  for example in  textbooks like  \cite{ CP} or \cite{Majidbook1}.  The information about the inner product used in the definition of  the Atiyah-Bott symplectic structure (or, equivalently, in the Chern-Simons action) is encoded in $r$ via the following compatibility requirement:

{\bf Definition:} An $r$-matrix is  compatible with a Chern-Simons action if it 
satisfies the CYBE  \eqref{cybe} and if its symmetric part is equal to the Casimir 
associated to the $\Ad$-invariant,
non-degenerate symmetric bilinear form  used in the Chern-Simons action.

In our case, the relevant Casimir operator for the `gravitational' bilinear form \eqref{gravinprod} is
\bea
\label{casimir}
K = 16\pi G (J_a\otimes P^a+P_a\otimes J^a).
\eea
A family of compatible $r$-matrices is given by \cite{Schroers, MeusburgerSchroers6}
\bea
\label{rmatsols}
 r= 32 \pi G \left   (P_a\otimes J^a + \epsilon_{abc}n^aJ^b\otimes J^c\right)           , \quad n_an^a=-\lambda,
\eea
where we use the metric \eqref{3dmetric} to lower and contract indices.

Two comments are in order here.  The first concerns the dependence of the solution on the  real vector $\bn=(n^0,n^1,n^2)$ which has to satisfy the given constraint but is otherwise arbitrary. Thus, for $\lambda < 0$, the vector $\bn$ is  any vector of length $\sqrt{-\lambda}$ in the Euclidean (hyperbolic) case, but is necessarily time-like in the Lorentzian (de Sitter) case.  For $\lambda =0$, $\bn$ vanishes in the Euclidean case but may be any light-like vector in the Lorentzian case. For $\lambda >0$, $\bn$ is space-like  in the Lorentzian (anti de-Sitter) case, while there is no  real solution in the Euclidean case. However, the Euclidean case with $\lambda >0$ (and hence $\Lambda >0$) is the only case where the  model space ($S^3$) and the local isometry group $SU(2)\times SU(2)$ are both compact, and the Chern-Simons theory is simply two copies of $SU(2)$ Chern-Simons theory, which  is extensively studied in the literature, see \cite{EMSS} for an early paper. I will not say  much about this case in the following, although it seems interesting and worthwhile to  relate the many results  about $SU(2)$ Chern-Simons theory  to the framework discussed here, and to interpret them in terms of 3d gravity. Presumably this would involve using a complex vector $\bn$ and imposing a suitable reality condition after  quantisation. 

The second comment concerns the non-uniqueness of the   solutions \eqref{rmatsols}. These solutions all   amount to equipping the Lie algebras $\glambda$ with the structure of a classical double, see \cite{CP,K-S} for general background  and \cite{MeusburgerSchroers6} for an explanation in the context of 3d gravity.  However,  other  $r$-matrices are known, which are also compatible with the  bilinear form \eqref{gravinprod} but which do not belong to the family \eqref{rmatsols}, see \cite{MeusburgerSchroers7} for examples and the forthcoming paper  \cite{OseiSchroers2} for a systematic discussion.  This gives rise to an ambiguity in the implementation of the Fock-Rosly prescription and the subsequent quantisation, but presumably leads to the same quantum theory. This issue has not been conclusively settled, and is also discussed in \cite{OseiSchroers2}.  One advantage of working with the $r$-matrices associated to classical doubles is that one may quantise by going to the associated quantum double. This is what we will review in the next section.

The Fock-Rosly Poisson structure on $\tilde {\cal P}$ is determined in  terms of a compatible   $r$-matrix. The formulae for the brackets are  explicit but lengthy, and we  refer  the reader to \cite{FockRosly}  or  \cite{AS}   for details. Some understanding of it can be gained from the observation,  made in  \cite{AMII}, that the Poisson brackets can be `decoupled' after a suitable coordinate change, and that, as a symplectic  manifold,    $\tilde {\cal P}$ is isomorphic    to a direct sum of  $g$ copies of the Heisenberg double of the Poisson-Lie group $\Glambda$ (with the Sklyanin Poisson-Lie structure defined by $r$) and the manifolds $ {\cal C}_{\mu_i \sigma_i}$, $i=1,\ldots n$ viewed as symplectic leaves of the dual Poisson-Lie group $\Glambda^*$:
\bea
\label{extendeddecoupled}
\tilde {\cal P}\simeq \text{Hei}(\Glambda) \times \ldots \times \text{Hei}(\Glambda) \times {\cal C}_{\mu_n \sigma_n}\times \ldots {\cal C}_{\mu_1 \sigma_1}.
\eea
The general definitions of the Sklyanin, Heisenberg double and dual Poisson structures  can be found in the  paper \cite{AMII} and also in the textbook \cite{CP} or  the lecture notes \cite{K-S}. We will give some further background in the next Section, but   here we note  that all of  these structures for the family of groups $\Glambda$ with the $r$-matrices \eqref{rmatsols} are explicitly given in  in \cite{MeusburgerSchroers6}. For example, in the case of vanishing cosmological constant (and $\bn$ vanishing), one finds  \cite{MeusburgerSchroers1,MeusburgerSchroers2}
\[
\text{Hei}(SL(\RR)\ltimes \RR^3)\simeq T^*(SL(2,\RR)\times SL(2,\RR)).
\]
In the Fock-Rosly description of the phase space \eqref{truephase} one still needs to  impose a constraint  in $\tilde {\cal P}$, and take a quotient. We will not pursue this here since we are mainly interested in the quantum theory.  Our approach to quantisation  is to  
quantise $\tilde {\cal P}$ first, and then to  take the quotient at the quantum level.


\section{Quantum groups and 3d quantum gravity}
\subsection{The combinatorial quantisation programme and associated quantum groups}
The task of constructing a quantum theory of 3d gravity in the Chern-Simons approach followed  here  is that  of quantising the Poisson algebra of functions on the physical phase space \eqref{truephase}, and of finding a  unitary,   irreducible  representation (UIR)  of  the quantised algebra. 
By `quantisation' of a Poisson manifold  $M$ we mean,  generally speaking, a deformation  $F_h(M)$ of the  algebra  of  functions on that manifold with a multiplication depending on a parameter $h$ in such a way that the commutator  of two elements in $F_h(M)$  to first order in that parameter equals the Poisson bracket of the classical limit of those elements \cite{CP}.  Details, for example the precise class of functions (${\cal C}^\infty$ or some algebraic subset),  depend on the Poisson manifold in question.

In the combinatorial approach, one simplifies this task by first quantising the extended phase space \eqref{extended}, and then imposing the reduction to \eqref{truephase} at the quantum level by  a suitable condition on the Hilbert space carrying  the UIR of the quantisation of \eqref{extended}. 
An important  advantage of the combinatorial approach is that one  really only needs to carry out the quantisation of  the building blocks entering the decomposition  of the extended phase space \eqref{extendeddecoupled}, and that these, in turn, can all be constructed from {\em one}  quantum group $H$ and its representations. 

The quantum group $H$  in question is the quantisation  of the so-called dual Poisson-Lie  group  $\Glambda^*$  of $\Glambda$ (with the Sklyanin Poisson-Lie defined by the $r$-matrix \eqref{rmatsols}).  This is explained in general terms in \cite{AGSI,AGSII} and in the particular case of semi-direct products like the Euclidean or Poincar\'e groups in \cite{MeusburgerSchroers2}.  It can be motivated as follows.

The dual Poisson-Lie group   $\Glambda^*$ is a non-linear analogue of the Kirillov-Kostant-Souriau (KKS)  Poisson structure on the dual $\glambda^*$  of the  Lie algebra $\glambda$  \cite{AMI,K-S}. Since the quantisation of the  KKS  structure  on $\glambda^*$ is  the universal enveloping algebra $U(\glambda)$, it is not surprising that the  quantisation of the Poisson algebra of $\Glambda^*$ is a deformation of  $U(\glambda)$. Thus we see already at this general level that  the quantum groups   $H$ are Hopf algebras obtained by  deforming  the  local isometry groups  $\Glambda$ (or more precisely, of  their group algebras). We therefore refer  to them as quantum isometry groups in the following. There is  further similarity between the canonical  Poisson structure on $\glambda^*$ and $\Glambda^*$: the symplectic leaves of  the former are co-adjoint orbits while the symplectic leaves of the latter are conjugacy classes in $\Glambda$ \cite{CP,K-S}. Given the non-degenerate bilinear pairing \eqref{gravinprod} on $\glambda$,  co-adjoint orbits may be thought of as adjoint orbits in $\glambda$,  and conjugacy classes in $\Glambda$ may  be thought of as non-linear deformations of these.

The irreducible representations of a Lie algebra can be obtained by quantising the KKS Poisson algebra and imposing the conditions which define the co-adjoint orbits  in terms of suitable Casimir operators. This analogy,  and the general comments  of the previous paragraph, go some way to motivating the result that the quantisation of the conjugacy classes ${\cal C}_{\mu_i\sigma_i}$ in the decomposition \eqref{extendeddecoupled}  gives UIRs $ V_{\mu_i\sigma_i}$ of the quantum group $H$ (with possible quantisation conditions on the labels $\mu_i,\sigma_i$). The quantisation of the classical Heisenberg double of $G_\lambda$ is the Heisenberg double of the Hopf algebra $H$ \cite{Majidbook1}. Its unique  irreducible representation, in the cases where they have been studied, is a quantum group analogue of the regular representation of a group, and we therefore denote it by  $\text{Reg}(H)$. We thus arrive at the following Hilbert space for the quantisation of  the extended phase space \eqref{extendeddecoupled}
\bea
\label{exthilbert}
\tilde {\cal H} = \text{Reg}(H)^g\otimes V_{\mu_n\sigma_n}\otimes \ldots V_{\mu_1 \sigma_1}.
\eea
This space is, by construction,  a (reducible) representation of the quantum group $H$. The
  Hilbert space for quantisation of the physical phase space \eqref{truephase} is the invariant part under this $H$-action \cite{AGSI,AGSII,AS}:
\bea
\label{truehilbert}
{\cal H}  = \text{Inv}_H (\tilde{\cal H}).
\eea

In order to carry out the combinatorial quantisation programme   in practice  one needs to  construct the  quantum group $H$ and to find  the representations appearing in \eqref{exthilbert}. The construction of  the quantum group $H$ is facilitated by the fact that the $r$-matrices \eqref{rmatsols} equip $\glambda$ with the structure of  a classical double of either $sl(2,\RR)$ (in the Lorentzian case) or $su(2)$ (in the Euclidean case) with suitable bialgebra structures, given in \cite{MeusburgerSchroers6}. Following the principle that the quantisation of the double is quantum double of the quantisation \cite{Drinfeld}, the family of quantum groups $H$ can thus easily be found. We list them  in Table~\ref{qisogroups}, which should be seen as a quantised and `gravitised' version of Table~\ref{isogroups} of the  classical isometry groups.  We will not give definitions or lists of generators and relations for any  of these quantum groups here, but refer to the standard textbooks \cite{CP,Majidbook1}.  However,  to gain some physical understanding  it is worth noting  that half the generators should be interpreted as rotation/boost generators and the other half as momentum generators. Thus, for example in the Lorentzian case of vanishing cosmological constant 
\bea
\label{double}
D(U(su(1,1)))=U(su(1,1))\ltimes \CC(SU(1,1)),
\eea 
as an algebra, where $\CC(SU(1,1))$ are complex-valued, smooth functions on $SU(1,1)$. The generators $J_a$ of $U(su(1,1)))$ are  simply  the rotation generator $J_0$ and the boost generators $J_1,J_2$ already encountered in \eqref{21alg}, while elements of  $\CC(SU(1,1))$ should be thought of as functions or  coordinates on the non-linear momentum space $SU(1,1)$, see \cite{sissatalk} for details and references, and also below for further remarks.  Finally, 
 the parameter $q$ appearing in the table is the one introduced at the beginning of this talk \eqref{qpara}. It combines all four physical parameters entering quantum gravity with a cosmological constant. 

\begin{table}[h!]
\baselineskip 16 pt
\begin{center}
 \begin{tabular}{|c|c|c|}
 \hline
   &    &   \\
 Cos. constant  & Euclidean ($c^2<0)$  & Lorentzian ($c^2>0$)  \\
   &    &   \\
 \hline
   &    &   \\
 $\Lambda  = 0$  & $D(U(su(2)))$ & $D(U(su(1,1)))$ \\
  &    &   \\
 \hline
   &    &   \\
 $\Lambda > 0$ & 
 $ D(U_q(su(2)))$, $q$ root of unity &
 $D(U_q(su(1,1)))$ $q\in  \RR$ \\
   &    &   \\
 \hline
   &    &   \\
 $\Lambda < 0$ & $D(U_q(su(2)))$, $q\in \RR$ & $D(U_q(sl(2,\RR)))$,  $q\in U(1)$ \\
   &    &   \\
 \hline
 \end{tabular}
\caption{Quantum isometry groups in 3d quantum gravity, $ q=e^{-\frac{\hbar G\sqrt{\Lambda}}{c}}$}
\label{qisogroups}
\end{center}
\end{table}

The combinatorial quantisation programme has been carried out to various degrees  of completeness in the different cases. For the Euclidean  case  with  vanishing cosmological constant, the  importance of the quantum double $D(U(su(2))$ was first pointed out in \cite{BM}, and the proof that it plays the role of the quantum isometry group $H$  in the combinatorial approach to Euclidean quantum gravity without cosmological constant was given in \cite{Schroers}. The  Lorentzian case was considered in \cite{BMS} and the general situation of Chern-Simons theory with certain semidirect product gauge groups was considered in \cite{MeusburgerSchroers2}. The situation where the classical gauge group is $SL(2,\CC)$ (i.e. Euclidean with $\Lambda <0$ or Lorentzian with $\Lambda >0$) was studied in \cite{BNR}, with the relevant quantum group already constructed in \cite{PodlesWoronowicz}. The Euclidean case with $\Lambda >0$ is  essentially the Turaev-Viro model. Finally, the very interesting Anti-de Sitter case (Lorentzian and $\Lambda >0$) has, unfortunately,  not received much attention  in the framework sketched here. 

\subsection{Non-commutative momentum addition, braiding and non-commutative spacetimes}

Having constructed the quantum groups  which control the construction of   3d quantum gravity according to the combinatorial scheme  it is natural to ask what one can learn  from them  about the physics of 3d quantum gravity. 

Formally, the role of the quantum  isometry groups listed in Table~\ref{qisogroups} is strictly auxiliary. The physical Hilbert space \eqref{truehilbert} is, by definition, invariant under the action of those quantum groups. Physical observables which act on this Hilbert space (see \cite{Meusburger6} for a discussion of classical examples) are not obviously related to the quantum isometry groups. As already mentioned (and discussed further in the Conclusion), the $r$-matrix used in the Fock-Rosly scheme, and hence the associated  quantum group, is not uniquely determined. Both of these observations suggest that the quantum groups in  Table~\ref{qisogroups} have only an indirect physical significance.

On the other hand, the quantum isometry groups, their representations and even their quantum $R$-matrices can be directly related to  physical properties of particles in 3d quantum gravity. We will illustrate this for the case of vanishing cosmological constant. In that case, the quantum doubles appearing in Table~\ref{qisogroups} are quantum doubles of the Lie groups $SU(2)$ in the Euclidean case  and  $SU(1,1)$ (which is isomorphic to $SL(2,\RR)$) in the Lorentzian case. These quantum doubles are semi-direct products as algebras  as shown in \eqref{double}, and have a representation theory which is very similar to those of the Euclidean and Poincar\'e groups \cite{BM,KM,KBM}. The only difference is that  the `mass shell' in momentum space which characterises  UIRs of the  Euclidean and Poincar\'e group  become conjugacy classes in the non-linear momentum  spaces ($SU(2)$  in the Euclidean case and $SU(1,1)$ in the Lorentzian case). Physically, this means that momenta are no longer vectors  but group elements of $SU(2)$ or $SU(1,1)$ and that momentum `addition' is implemented by group multiplication in  $SU(2)$ or $SU(1,1)$  instead of vector addition. These non-linear and non-commutative  properties of momentum addition for gravitating particle reflect the  use of holonomies for characterising particle properties, as y used in  early papers on 3d gravity \cite{DJtH,Carlipscattering}. We can even see it in the simplest non-trivial  example  of  3d  spacetime, namely the cone shown in Fig.~\ref{conedeflection}. The spacetime is fully characterised by the deficit angle $\mu$, which is the mass of the particle in units of the Planck mass $1/8\pi G$. However, the angular nature of this parameter   fits very well into the picture of $SU(1,1)$-valued momenta: we simply  think of $\mu$ as a rotation, i.e. a particular element of $SU(1,1)$.

A closely related property of gravitating particles is their scattering, as analysed in some of the early papers on 3d quantum gravity \cite{DeserJackiw2,tHooftscattering}. It turns out that the $S$-matrix for  the scattering of two massive and spinning particles  can also be interpreted in terms of quantum groups and the sort of topological interactions discussed in Sect.~\ref{intro}. As shown in \cite{BMS}, the $S$-matrix is   naturally related to the $R$-matrix of the quantum double $D(U(su(1,1)))$.

Finally, the curved   and non-abelian nature of the momentum manifold suggests that naturally defined positions coordinates (which should generate translations on momentum space) should be non-commutative. One can argue this more formally by demanding that momentum and position algebras should be dual as Hopf algebras, leading to the family of Hopf algebras shown in Table~\ref{qmompos}. A particular, and much studied example  is the `spin spacetime' with generators $X_0,X_1,X_2$ and commutation relations 
\bea
[X_a,X_b]=\ell_P \epsilon_{abc}X^c,
\eea
where $\ell_P= 8\pi \hbar G$ is the Planck length in 3d gravity, and both the Euclidean and Lorentzian interpretation apply. This non-commutativity of positions was already considered  in \cite{tHooftdiscrete} and \cite{MatWell}, and appears naturally in the quantum group theoretical framework considered here. It can also be derived in other approaches, namely in a path integral  for particles  where gravitational field  degrees of freedom have  been integrated out  \cite{FreidelLivine} or  in a coset construction \cite{JoungMouradNoui}, which is analogous to the way  the classical  spacetimes \eqref{hypersur} can be obtained as  homogeneous spaces  of the classical isometry groups $\Glambda$.  Finally,  the  role of the quantum double $D(SU(2))$ as a quantum isomtetry group  of  the 3d (Euclidean) was noted in \cite{BatistaMajid}, where the latter was studied from the point of view of non-commutative differential geometry. 

It is interesting that  physical arguments, path integrals, coset constructions and general quantum group theoretical considerations all lead to the same non-commutative spacetimes.  One way of exploring the physical significance of this non-commutativity is to study representations of the quantum doubles in Table~\ref{qisogroups} in position space. The requires Fourier-transforming the 
usual formulation of the representations in momentum space, in analogy to the way the UIRs of the Poincar\'e group can be Fourier transformed into  the solution space of the  familiar wave equations of  relativistic physics (Klein-Gordon, Dirac, Maxwell etc).  This was carried out  for $D(SU(2))$ in \cite{MajidSchroers} and is considered for the Lorentzian case in \cite{SchroersWilhelm}. 
 
\begin{table}[h!]
\baselineskip 16 pt
\begin{center}
 \begin{tabular}{|c|c|c|}
 \hline
   &    &   \\
 Cos. const.   & Euclidean ($c^2<0)$  & Lorentzian ($c^2>0$)  \\
   &    &   \\
 \hline
   &    &   \\
 $\Lambda  = 0$  & $\CC(SU(2)) \;/\; U(su(2))$ & $\CC(SU(1,1)))\;/\; U(su(1,1))$ \\
  &    &   \\
 \hline
   &    &   \\
 $\Lambda > 0$ & 
 $ \CC_q(SU(2))\;/\;  U_q(su(2))$, $q$ root of unity &
 $\CC_q(SU(1,1))\;/\; U_q(su(1,1))$ $q\in  \RR$ \\
   &    &   \\
 \hline
   &    &   \\
 $\Lambda < 0$ & $\CC_q(SU(2))\;/\; U_q(su(2))$, $q\in \RR$ & $\CC_q(SL(2,\RR))\;/\; U_q(sl(2,\RR))$,  $q\in U(1)$ \\
   &    &   \\
 \hline
 \end{tabular}
\caption{Momentum/position algebras  in 3d quantum gravity,   $ q=e^{-\frac{\hbar G\sqrt{\Lambda}}{c}}$}
\label{qmompos}
\end{center}
\end{table}

\section{Outlook and conclusion}

We have seen that the combinatorial quantisation of the    Chern-Simons formulation of 3d gravity gives  a unified picture of the various regimes
of 3d gravity, with the physical parameters $c,\Lambda, G $ and $\hbar$ entering as deformation parameters in distinctive ways.  Quantum groups naturally replace the classical isometry groups in this approach to 3d quantum gravity, and non-commutative spacetimes replace the classical model spacetimes. In general, the relation between the quantum isometry groups and the physical Hilbert space of 3d quantum gravity is a formal one, but we have seen that aspects of the quantum isometry groups like the non-commutative momentum addition and the braiding via the quantum $R$-matrix have a direct physical interpretation. It is worth noting that it is possible to take a Galilean limit $c\rightarrow \infty$ in the framework discussed here \cite{PapageorgiouSchroers1,PapageorgiouSchroers2}, and that the non-commutative quantum space is the Moyal plane in that case, with a  time-dependent non-commutativity of the spatial coordinates.

In order to clarify the physical interpretation of quantum isometry  groups  and the associated   non-commutative spacetimes it may be useful to consider universes with a boundary instead of the spatially compact universes considered in this talk.  The treatment of boundaries  in the classical theory is discussed in \cite{Matschull2,MeusburgerSchroers4,MeusburgerSchroers5} but a general treatment of the quantisation has not been given. Another  approach would be to work directly on the physical phase space as in \cite{Meusburger6,MeusburgerSchonfeld}, and to attempt the  quantisation there. 

Other quantum groups than quantum doubles  have been discussed in relation to 3d quantum gravity, notably bicrossproducts or  $\kappa$-Poincar\'e algebras which were originally introduce in 4d \cite{LNRT,LNR,MajidRuegg}. As shown in \cite{MeusburgerSchroers7},   the $\kappa$-Poincar\'e algebra with the usual time-like deformation parameter is {\em not} compatible with 3d gravity in the combinatorial framework. On the other hand, $\kappa$-Poincar\'e algebras with space-like deformation parameters are possible.  This and other quantisation ambiguities of 3d quantum gravity are discussed in the forthcoming paper \cite{OseiSchroers2}.

\end{document}